\documentclass[aps,prb,twocolumn,epsf,epsfig,amsmath]{revtex4}

\usepackage{latexsym}
\usepackage{hyperref}
\usepackage{times}
\usepackage{graphicx}
\usepackage{amsmath}
\usepackage{braket}
\usepackage{dcolumn}
\usepackage{bm}
\usepackage{textcomp}
\usepackage{color}
\usepackage{amssymb}
\usepackage{xcolor}
\usepackage{easyReview}
\usepackage{mathdots}
\usepackage{float}
\usepackage{lineno}
\usepackage{todonotes}
\usepackage{array}
\usepackage[export]{adjustbox}
\newcommand{\beq}{\begin{equation}}
\newcommand{\eeq}{\end{equation}}

\newcommand{\bb}[1]{{{\mathbf #1}}}

\begin{document}
\title{Proof of a Stable Fixed Point for Strongly Correlated Electron Matter}
\author{Jinchao Zhao$^1$, Gabriele La Nave$^2$, and Philip W. Phillips$^1$}

\affiliation{$^1$Department of Physics and Institute for Condensed Matter Theory,
University of Illinois
1110 W. Green Street, Urbana, IL 61801, U.S.A.}
\affiliation{$^2$Department of Mathematics, University of Illinois, Urbana, Il. 61801}

\begin{abstract}
We establish the Hatsugai-Kohmoto model as a stable quartic fixed point (distinct from Wilson-Fisher) by computing the $\beta-$function  in the presence of perturbing local interactions.  In vicinity of the half-filled doped
Mott state, the $\beta-$function vanishes for all local interactions regardless of their sign. The only flow away from the HK model is through the superconducting channel which lifts the spin degeneracy as does any ordering tendency.  The superconducting instability is identical to that established previously\cite{nat1}.  A corollary of this work is that Hubbard repulsive interactions flow into the HK stable fixed point in the vicinity of half-filling.  
Consequently, although the HK model has all-to-all interactions, nothing local destroys it. The consilience with Hubbard arises because both models break the $Z_2$ symmetry on a Fermi surface, the HK model being the simplest to do so.  Indeed, the simplicity of the HK model belies its robustness and generality. 
\end{abstract}
\date{March 2023}
\maketitle

\section{Introduction}

\begin{figure}[h]
\includegraphics[scale=.45]{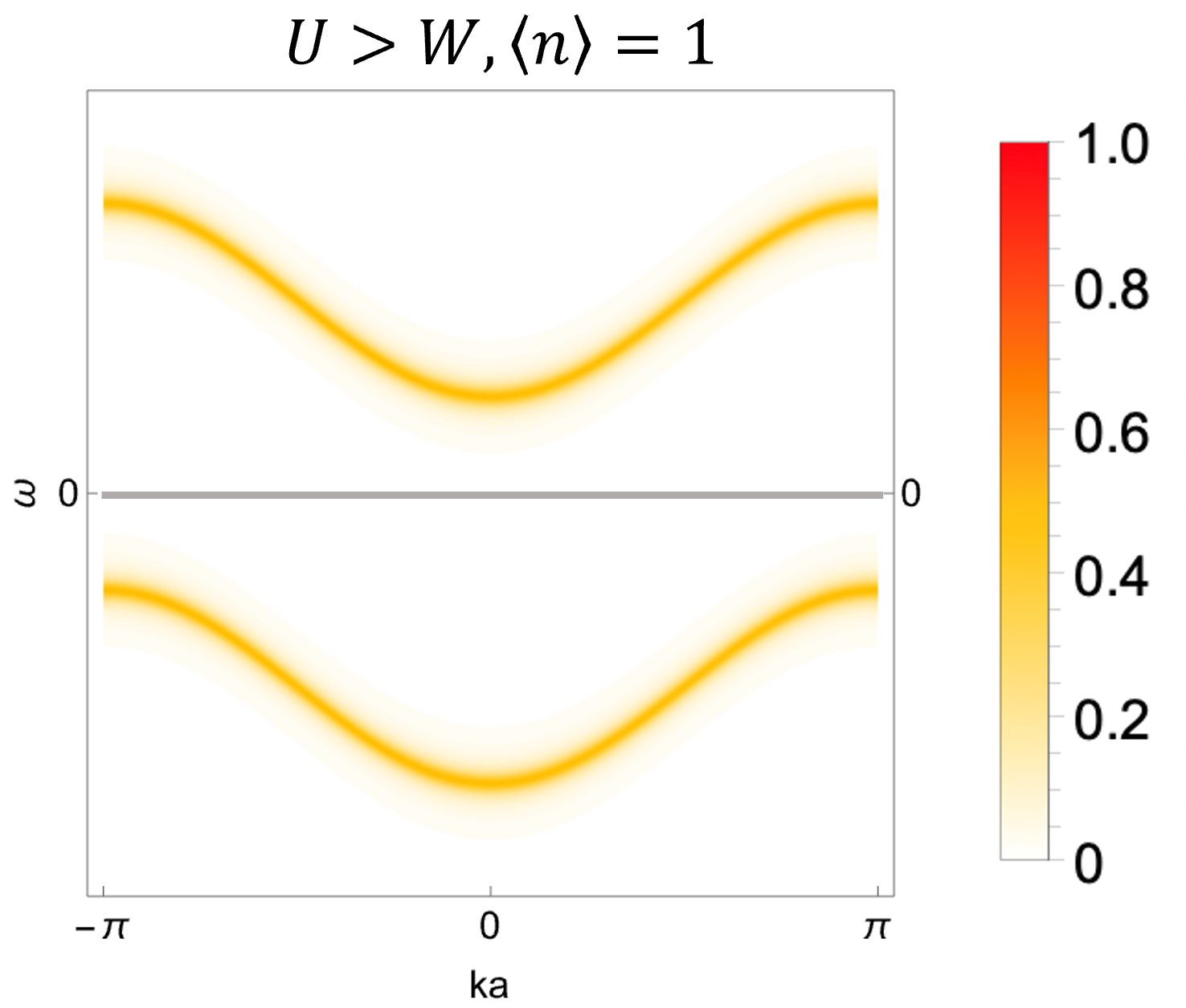}
\includegraphics[scale=.45]{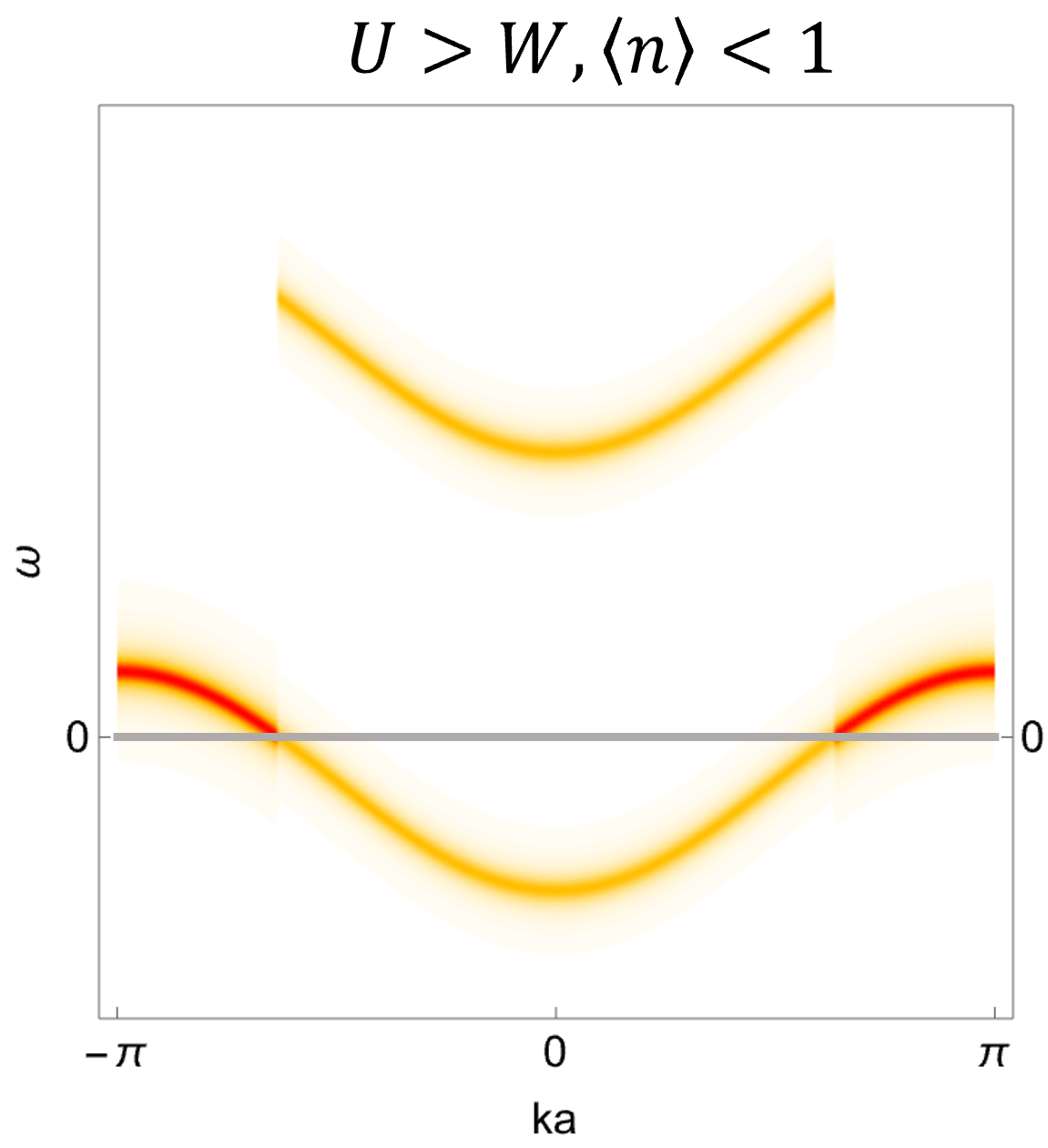}
\includegraphics[scale=.45]{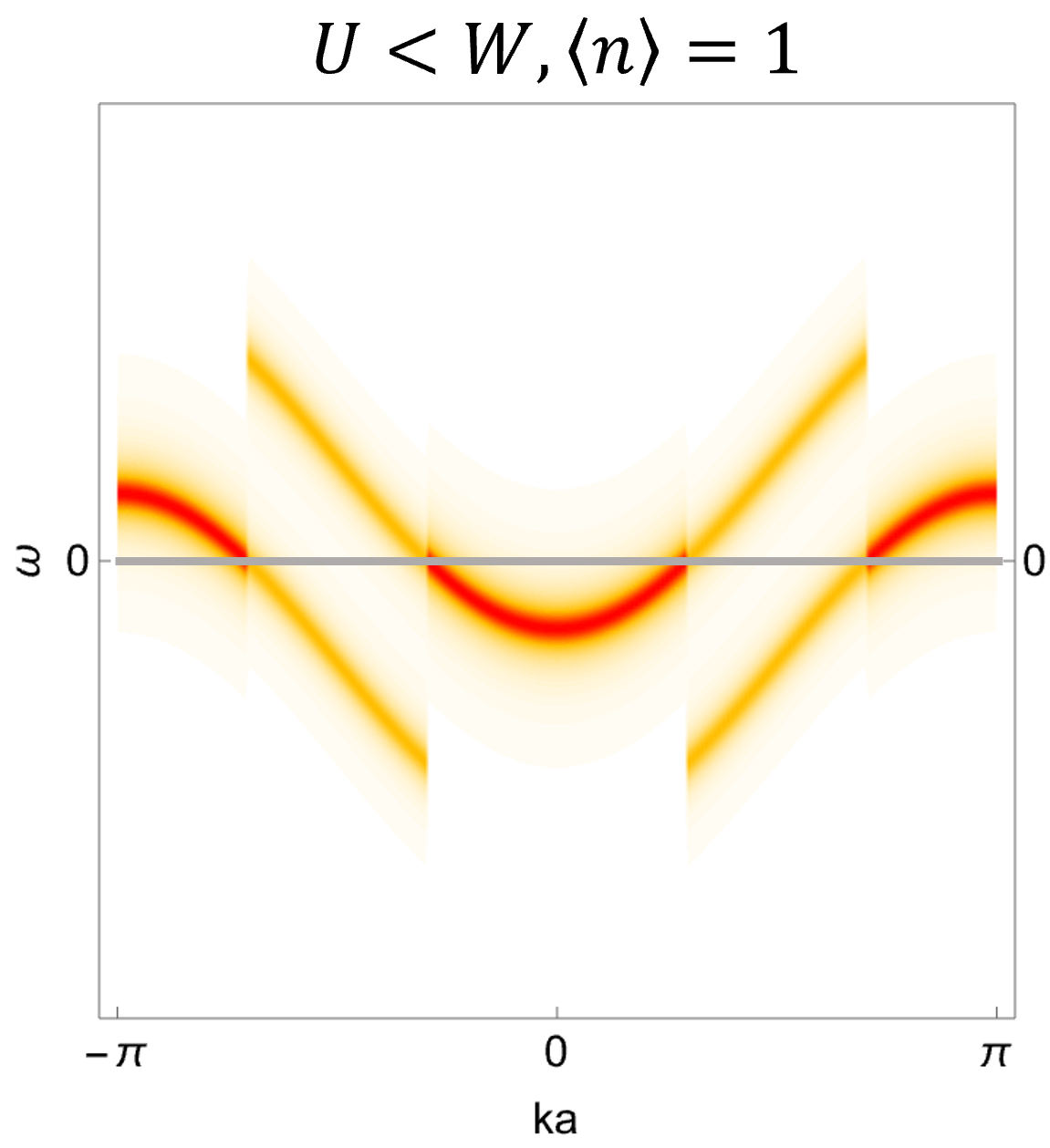}
\caption{The spectral functions of the HK model at different fillings and interaction strength. (\textbf{a}) Mott insulating state, with $U>W$ and $\braket{n}=1$. (\textbf{b}) Strongly repulsive Mott metal, with $U>W$ and $\braket{n}<1$. (\textbf{c}) Weakly repulsive Mott metal, with $U<W$ and $\braket{n}=1$.}
\label{spectral}
\end{figure}

\begin{figure}[ht]
\includegraphics[width=\columnwidth]{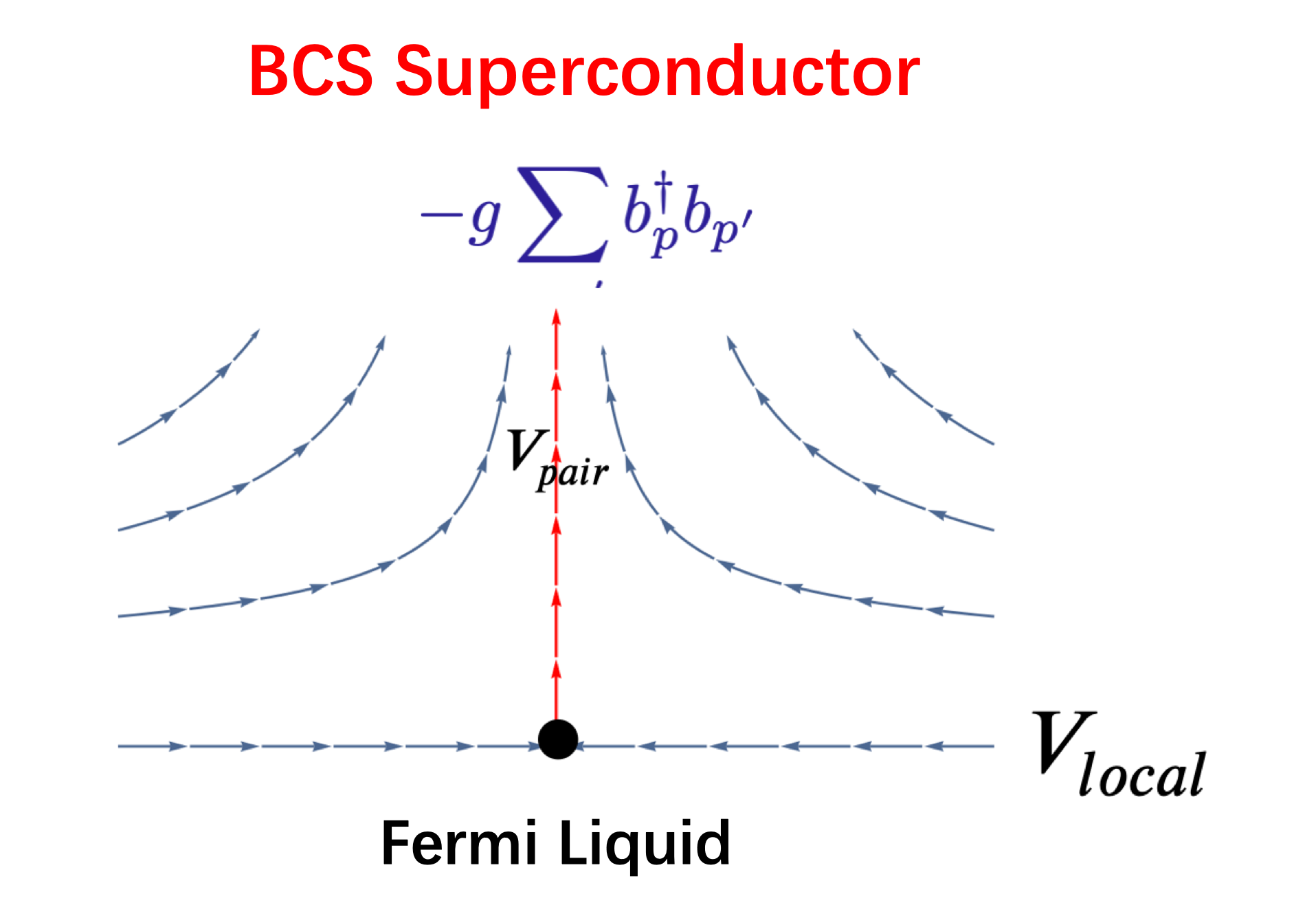}
\caption{Perturbative low diagram for interactions in a Fermi liquid.  Short-range interactions regardless of their sign do nothing.  Pairing leads to a flow to strong coupling and the ultimate destruction of the Fermi liquid and the onset of a superconducting state.  The nature of the superconducting state cannot be established based on perturbative arguments but requires BCS theory.}
\label{flflow}
\end{figure}

Proving Mott's claim, in any dimension we care about ($d\ge 2$), that strong electron correlation opens a gap in a half-filled band without changing the size of the Brillouin zone still stands as a key unsolved problem in theoretical physics.   Demonstrating that the spectral function contains no states in momentum space that cross the chemical potential would suffice as proof of Mott's claim.  However, the inherent problem is that the model employed in this context, namely the Hubbard model, contains strong on-site repulsion in real space, thereby preventing any exact statement about the corresponding spectral function in momentum space. There is of course an easy way around this problem: add to the non-interacting band model an energy penalty, $U$, 
 \beq
 H=\sum_{\bb p\sigma} \xi_{\bb p\sigma}n_{\bb p\sigma}+U\sum_{\bb p} n_{\bb p\uparrow}n_{\bb p\downarrow},
 \label{eq1}
 \eeq
 whenever two electrons doubly occupy the same momentum state. In the above $\xi_{\bb p\sigma}$ is the non-interacting band dispersion and $n_{\bb p\sigma}$ is the occupancy.   Since such an interaction does not mix the momenta, a gap must open in the spectrum should $U$ exceed the bandwidth as illustrated in Fig. \ref{spectral}.  This is precisely the Hatsugai-Kohmoto model\cite{hk} which was introduced in 1992 but attracted essentially no attention until\cite{nat11,nat12,nat3,nat4,nat5,nat6,nat7,nat8,nat9,nat10,natfr}  our demonstration\cite{nat1,nat2} that this model offers an exact way of treating the Cooper instability in a doped Mott insulator.   Despite this utility, the HK model faces an uphill battle to replace the ingrained Hubbard model as the knee-jerk response to the strong-correlation problem with local-in space interactions.   In bridging the gap from the HK to the Hubbard model, three questions arise.  1) Does the HK model remain stable to local-in-space interactions?  If it does, then a correspondence with the Hubbard model can be established.  2)  Does the resilience to local-in-space interactions give rise to a stable fixed point?  3) What about the obvious spin degeneracy that arises from the singly occupied states in the spectrum?  Are there ordering instabilities at $T=0$ that lift the degeneracy?  
It is these three questions that we answer in this paper.  Briefly, we construct the $\beta-$function explicitly and demonstrate that  the answer to all three leading questions is yes.   The degeneracy is lifted spontaneously by superconductivity as in a Fermi liquid.  Central to our construction is the leading lesson from Fermi liquid theory.  As Fermi liquids admit a purely local description in momentum space, their
eigenstates are indexed simply by momentum and spin. Consequently, in real
space, Fermi liquids exhibit long-range entanglement. It is precisely this long-
range entanglement that makes them impervious to any local interaction, that is, short-range  Coulomb interactions.  
The renormalization group analysis of a Fermi liquid shows distinctly\cite{polchinski,shankar} that  should the sign of the interaction be reversed and the electrons reside
 on opposites of the Fermi surface, a superconducting instability arises.  The added stipulation of the electrons residing on opposite sides of the Fermi surface changes the scaling dimension of the generic 4-fermion interaction from being irrelevant to marginal.  Also of note is that in terms of the pair annihilation operator, $b_k=c_{k\uparrow}c_{k\downarrow}$, the Cooper pairing interaction 
 \beq
 V_{\rm pair}=-g\sum_{\bb k,\bb k'}b_{\bb k}^\dagger b_{\bb k'},
 \eeq
 is completely non-local in real space as it involves an interaction between all pairs of electrons regardless of their separation.  Such non-locality does not make the BCS model unphysical because this interaction is the only relevant perturbation that  destroys a Fermi liquid as indicated by the 1-loop flow equation,
 \beq
 \frac{d g}{dt}=\frac{g^2}{4\pi},
 \eeq
which  exhibits a breakdown at a finite energy scale (the transition temperature) signaling the onset of pairing and the eventual non-perturbative physics of the BCS superconducting state.   All of this is summarized in the flow diagram in Fig. (\ref{flflow}):  short-range interactions ($V_{\rm local}$) regardless of their sign flow to the Fermi liquid state while the Cooper term, ($V_{\rm pair}$), inherently non-local in real space, flows to strong coupling.   Once again, it is the real-space long-range entanglement of a Fermi liquid that accounts for its resilience to any short-range interaction.  
 
 As it is ultimately the conservation of the charge currents in momentum space, $n_{\bb p\sigma}$,  that is at play here, it is natural to rewrite the HK model as $\sum_p h_p$ with $h_p$ implicitly defined from Eq. (1).  From this, it is clear that the HK model has a momentum-diagonal structure of a Fermi liquid.  However, unlike a Fermi liquid, this model describes a Mott insulator\cite{nat1,nat2,hk} as shown in Fig. (\ref{spectral}).  While it is natural to criticize this model as being unphysical because of the non-local in real space interactions, non-locality by itself does not dismiss a model from being physically relevant as $V_{\rm pair}$ in a Fermi liquid is non-local but nonetheless is the only relevant pairing term that leads to the BCS instability. 
 The real question is: Do local interactions lead to flow away from the Hatsugai-Kohmoto model?  That is, does the Mott insulator that HK describes represent a fixed point in direct analogy with the stability of a Fermi liquid to local interactions?  We show here that the answer to this question is a resounding yes; HK is a stable fixed point.  Even adding local terms of the Hubbard kind does nothing.

Analogies with Fermi liquid aside, a more fundamental reason is operative for the resilience of the HK model to local perturbations.  Haldane and Anderson\cite{Haldane} argued that Fermi liquids contain an added $Z_2$ symmetry on the Fermi surface that amounts to a particle-hole interchange for one of the spin species.   Operationally, if only doubly occupied or empty sectors are present, then adding or removing an electron of either spin is kinematically equivalent.  However, removing an electron from a singly occupied k-state can only be done in one way, thereby breaking the $Z_2$ symmetry.  Note even in the Hubbard model, the on-site repulsion creates single occupancy in real space which will give rise to single occupancy also in momentum space.   HK suffices as it is the simplest term that does so\cite{nat2}.  As long as this symmetry is already broken, adding new interactions which also break this symmetry yields no relevant physics as far as Wilsonian renormalization is concerned. 

We carry out the full renormalization group analysis for the HK model and show that as in a Fermi liquid, no local perturbation destroys the HK model, thereby defining a stable fixed point.  The only perturbation that destroys HK physics is $V_{\rm pair}$ as in a Fermi liquid.  We conclude then that the HK model is more than a toy model.  Rather it represents a stable fixed point for a model non-Fermi liquid that describes a Mott insulator.  It is a new fixed point in quantum matter that describes a doped Mott insulator.  The superconducting state that ensues\cite{nat1} is the BCS analogue for a doped Mott insulator.  

\section{RG approach for $d\ge 2$ HK model: Tree level analysis}

Our goal is to establish the stability of the HK model to weak local interactions.  What is central to the HK model are the two filling surfaces, the lower filling surface (L-surface) separates singly- and empty states, and the upper filling surface (U-surface) separates the doubly- and singly- occupancy. The key argument of the RG process is that at weak coupling, only modes around these filling surfaces  participate. At different filling and interaction strengths, the number of filling surfaces varies from zero (Mott insulating state, Fig.\ref{spectral}(a)), to one (strongly repulsive Mott metal, Fig.\ref{spectral}(b)) or two (weakly repulsive Mott metal, Fig.\ref{spectral}(c)). For simplicity, we work with $d=2$ and rotationally invariant filling surfaces. The higher dimensional result can always be achieved by adding rotational degrees of freedom and the relevance of interactions that define the fixed point is not changed.
\subsection{L-surface}

We start with the analysis of the setup with only one spherical filling surface of which the particle occupancy is singly- or empty on either side (L-surface). This analysis follows the method Weinberg\cite{weinberg} and Shankar\cite{shankar} though we use the notation of Polchinski\cite{polchinski}.  We linearize the dispersion around the L-surface with radius $K_L$ to 
\begin{equation}
    E\left(\bb K =\bb n (K_L+k)\right)=v_Lk,
\end{equation}
where $\bb{n}=\frac{\bb{K}}{|\bb{K}|}$ is the unit vector in the direction of $\bb{K}$ and $v_L$ is the isotropic L-surface velocity.  Note linearization restricts our analysis to the vicinity of metallic state where Mott physics is relevant.  Near the bottom of the band, our analysis fails as the band dispersion is inherently quadratic, and weak coupling physics obtains.
We write the zero temperature partition function, 
\begin{align}
&Z=\int\mathcal{D}[c,\bar{c}]e^{-S_0},\label{eq:LFZ}\\
\begin{split}
    &S_0=\int_\Lambda \frac{d^d\bb K}{(2\pi)^d}\int_{-\infty}^{\infty} d\omega\\
    &\left[\sum_{\sigma}\bar{c}_{\bb K\sigma}(i\omega-v_Lk)c_{ \bb K\sigma}+U\bar{c}_{\bb K\uparrow}\bar{c}_{\bb K\downarrow}{c}_{\bb K\downarrow}{c}_{\bb K\uparrow}\right].
\end{split}
\end{align}
The integral over momentum is confined in a thin shell around the filling surface, with distance $\Lambda$ as the cut-off. 
The partition function factorizes at each momentum $K$, in exactly the same ways as a Fermi liquid, however, mixing the up and down spins at the same momentum. 
After integrating out the fast modes living within $s\Lambda <|k|<\Lambda$, we perform the rescaling of the variables and fields:
\begin{align}
    k'&=s^{-1}k,\\
    \omega'&=s^{-1}\omega,\\
    c'_{\bb K'\sigma }&=s^{3/2}c_{\bb K\sigma },\\
    U'&=s^{-2}U,
\label{eq:scaling}
\end{align}
to make the partition function invariant up to a constant. It is worth noting that the HK repulsion has a scaling dimension of $-2$ and hence is strongly relevant. This relevant term suppresses the contribution of spectral weight from the other band that is $U$ away from the filling surface.

Now we consider the effect of perturbations on this fixed point.
First, consider perturbations that are quadratic in the fields,
\beq
    \delta S^{(2)}=\int_\Lambda \frac{d^d\bb K}{(2\pi)^d}\int_{-\infty}^{\infty} d\omega\sum_\sigma\mu(\bb K)\bar{c}_{\bb K\sigma}c_{\bb K\sigma }.
\eeq
This action separates into slow and fast pieces, and the effect of integrating out the fast modes produces a constant. After rescaling the momenta and the fields, we have
\begin{equation}
    \mu'(k')=s^{-1}\mu(k).
\end{equation}
This relevant term is the chemical potential, which should be included in the action kinetic term. As a result, the location of the fixed point definitely depends on the filling of the system. We shall adjust the position of the filling surface according to the chemical potential so as to make the system truly fixed.

Next, we consider the quartic interaction in the most general form,
\begin{equation}
\begin{split}
    \delta S_4&=\int_{K}\int_{\omega} \bar{c}_4(\tau)\bar{c}_3(\tau)c_2(\tau)c_1(\tau)u(4,3,2,1),\\
    \bar{c}_i&\equiv\bar{c}_{\bb K_i \sigma_i},\\
    c_i&\equiv c_{\bb K_i \sigma_i},\\
    u&(4,3,2,1)=u(\bb K_4 \sigma_4,\bb K_3 \sigma_3,\bb K_2 \sigma_2,\bb K_1 \sigma_1),\\
    \int_{K}&\equiv \prod_{i=1}^4\int_\Lambda d\bb {K}_i \int d\Omega_i \delta(\bb {K}_1 +\bb {K}_2 -\bb {K}_3 -\bb {K}_4 ),\\
    \int_{\omega}&\equiv \prod_{i=1}^4\int_{-\infty}^{\infty}d\omega_i\delta(\omega_1+\omega_2-\omega_3-\omega_4).
\end{split}
\end{equation}
The delta functions put constraints on the integral region of momentum and frequency. The delta function on frequency could be easily rescaled as $\delta(\omega_1'+\omega_2'-\omega_3'-\omega_4')=s\delta(\omega_1+\omega_2-\omega_3-\omega_4)$ since the integral over frequency extends to infinity. The delta function on momentum, however, has a different scaling behavior as pointed out by Polchinski\cite{polchinski}. The distance from the filling surface only gives a contribution proportional to the cutoff $\Lambda$ which is a negligible contribution compared with the filling momentum $K_L: \delta(\bb {K}_1 +\bb {K}_2 -\bb {K}_3 -\bb {K}_4 )\approx\delta(K_L(\bb{n_1} +\bb n_2-\bb n_3-\bb{n_4})+O(\Lambda))$. When the momenta all point in different directions, the first term in the delta function dominates and the delta function does not scale upon RG.
Following the same argument by Polchinski\cite{polchinski}, this quartic operator is thus irrelevant at the tree level,
\begin{equation}
    u'(4',3',2',1')=su(4,3,2,1).
\end{equation}
It is then easy to see that any further interactions are even more irrelevant.
This power-counting tree-level analysis already rules out the general short-range interactions from being relevant. Upon decreasing the energy scale, we find that the coupling becomes weaker and weaker while the HK repulsion gets stronger and stronger. Consequently, Mott physics captured by the filling surface is stable under local interactions.

There is an important subtlety in the kinematics and as a result, our treatment of the delta function is not always valid. The first-term contribution to the delta function could be set exactly to zero by putting an additional constraint on the momentum directions. The second term which is proportional to $\Lambda$ will be renormalized and generates a factor of $s$. The rescaling of these direction-constrained interactions are
\begin{equation}
    u'(4',3',2',1')=s^{0}u(4,3,2,1).
\end{equation}
Performing a Taylor expansion in $k$ and $\omega$ and comparing coefficients of separate powers, we conclude that the leading term, with no dependence on either variable, is marginal, while all the rest are irrelevant.
Because there are only two degrees of freedom in momentum, both of them are non-local by definition. 
Consequently, to reach a definitive conclusion, we must proceed to  the 1-loop level to determine whether they are marginally irrelevant or marginally relevant, or truly marginal. This calculation will be performed in the next section.

\subsection{U-surface}

In the case that the filling surface is the occupancy boundary of the doubly and singly occupied regions, we have to perform an extra particle-hole transform before we write down the partition function in the path integral language so as to obtain the correct linearized dispersion around the U-surface with radius $K_U$,
\begin{equation}
    E\left(\bb K =\bb n (K_U+k)\right)=v_Uk.
\end{equation}
This  particle-hole asymmetry reflects the broken $Z_2$ symmetry of Mott physics. The zero temperature partition function then becomes
\begin{align}
&Z=\int\mathcal{D}[c,\bar{c}]e^{-S_0},\label{eq:UFZ}\\
\begin{split}
    &S_0=\int_\Lambda \frac{d^d\bb K}{(2\pi)^d}\int_{-\infty}^{\infty} d\omega\\
    &\left[\sum_{\sigma}\bar{c}_{\bb K\sigma}(i\omega-v_Uk)c_{ \bb K\sigma}+U\bar{c}_{\bb K\uparrow}\bar{c}_{\bb K\downarrow}{c}_{\bb K\downarrow}{c}_{\bb K\uparrow}\right].
\end{split}
\end{align}
choosing the same setup with cut-off $\Lambda$ results in the same scaling rule for the variables and fields as in Eq.(\ref{eq:scaling}). The identical analysis on quadratic and quartic perturbations thus reoccurs and we have the same irrelevant tree-level behavior around this U-surface fixed point.

\subsection{2 Filling Surfaces}

In the weakly repulsive case, the 2 occupancy boundaries(L-surface and U-surface) coexist. In order to discuss low-energy physics, we have to include modes around both filling surfaces. Due to the factorizability of the partition function in momentum space, we can safely achieve the bare partition function by the product of Eq.(\ref{eq:LFZ}) and Eq.(\ref{eq:UFZ}). By setting the energy scale $\Lambda$ around both filling surfaces to be the same, we arrive at the same scaling of the variables and fields as in Eq.(\ref{eq:scaling}). 
In conclusion, regardless of the number of filling surfaces, local interactions are always irrelevant at the tree level and hence do not modify the fixed point defined by the filling surfaces. We will move on to see the effect of marginal quartic interactions on these fixed points.

\section{The perturbative expansion for 1-loop level corrections}

With the tree-level analysis in hand, we have already ruled out the local part of any perturbations. It is interesting to see how we can obtain a collective effect such as superconductivity in a low-energy theory.

The RG process is carried out by integrating the fast modes and rescaling the slow modes and the associated variables to keep the partition function unchanged. Besides the terms that only contain slow modes(tree-level result), we also need to include the terms that have both slow modes and fast modes and add their contribution to the scaling equations. This process is mathematically equivalent to calculating multipoint correlation functions with interactions. In the HK model, the correlation functions could be calculated using perturbative expansion for weak coupling as demonstrated in the Appendix. 

\begin{figure}[ht]
\centering
\includegraphics[scale=.22]{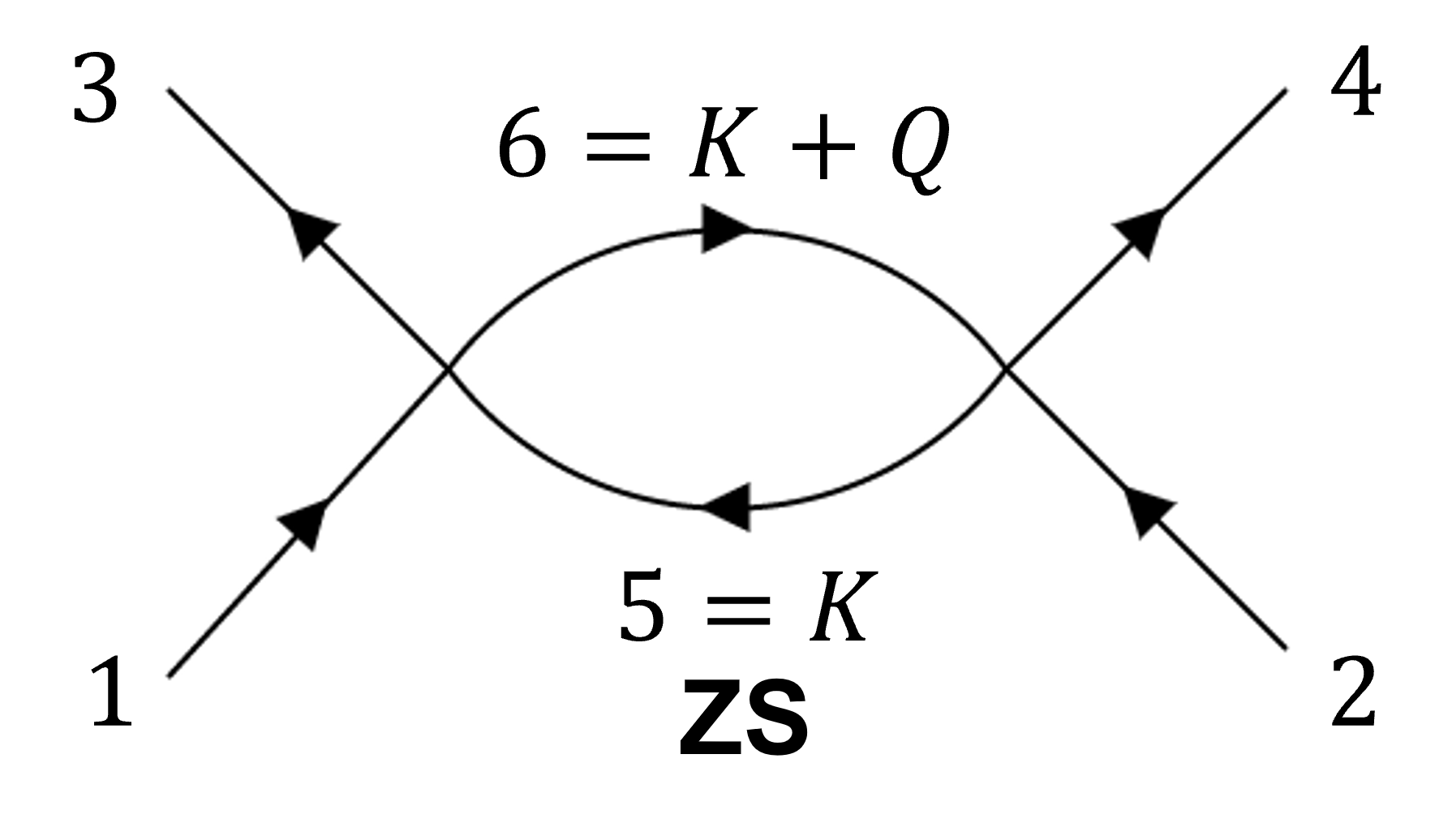}
\includegraphics[scale=.22]{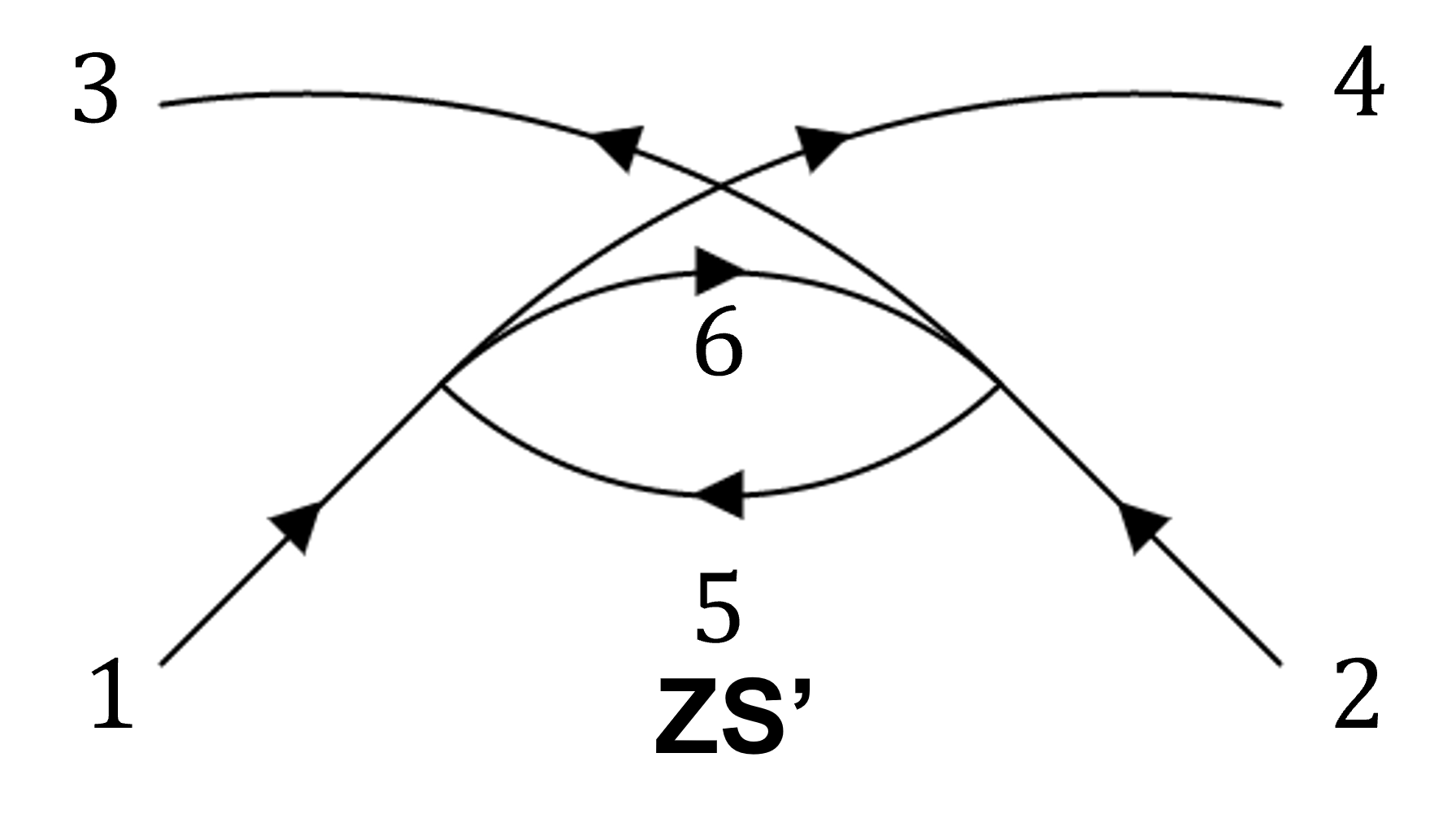}
\includegraphics[scale=.22]{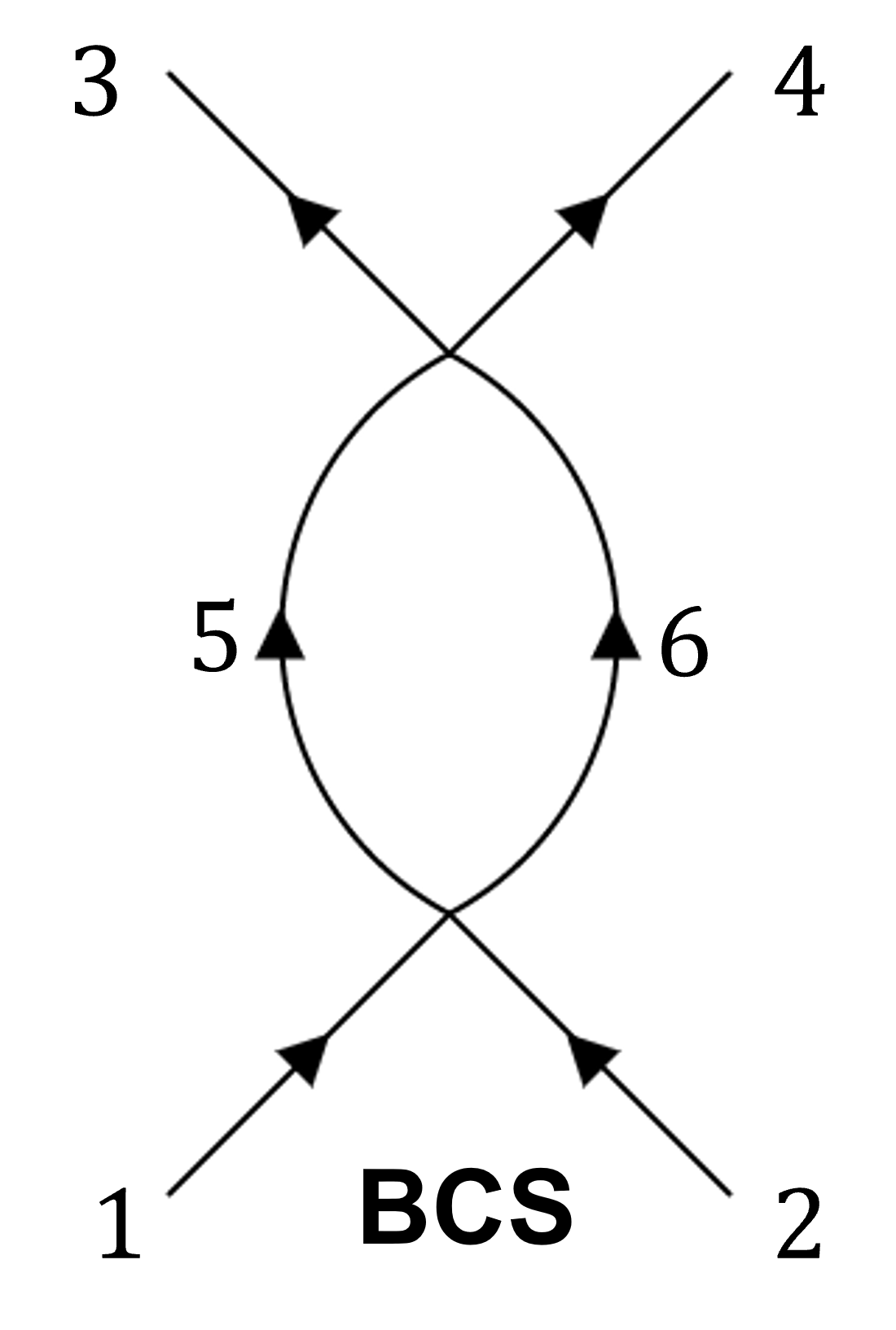}
\caption{The one-loop graphs for $\beta(u)$ for quartic interactions. The loop momenta lie in the shell of width $d\Lambda$ being eliminated. The external frequencies being all zero, the loop frequencies are (a) equal for ZS, (b) equal for ZS', and (c) equal and opposite for the BCS graph.}
\label{feynmandiagram}
\end{figure} 

The increment in $u(4,3,2,1)$ 
\begin{widetext}
\begin{equation}
\begin{split}
    du(4,3,2,1)=&\int u(6,3,5,1)u(4,5,2,6)G(5)G(6)\delta(3+6-1-5)d5d6\\
    &-\int u(6,4,5,1)u(3,5,2,6)G(5)G(6)\delta(6+4-1-5)d5d6\\
    &-\frac{1}{2}\int u(6,5,2,1)u(4,3,6,5)G(5)G(6)\delta(5+6-1-2)d5d6.
\end{split}
\label{eq:diags}
\end{equation}
\end{widetext}
is given by 3 diagrams as mentioned by Shankar, and we will follow the nomenclature to call them ZS, ZS' and BCS diagrams respectively.

We examine first the tree-level marginal interactions. We define the momentum components on the filling surface of $K_i$ as $K_i^F$. Then a common property of these marginal interactions is that the sum of the incoming and outgoing $K_i^F$s is zero $\bb {K}_1^F +\bb {K}_2^F -\bb {K}_3^F -\bb {K}_4^F =0$. The delta function on momentum thus scales as $s^{-1}$ and gives the marginal power counting. This property reduces the marginal interactions into two families:  1) forward scattering and 2) superconductivity or the Cooper channel.

\section{Forward scatterings at 1-loop}

The forward scatterings are defined by a non-vanishing $\bb{P}=\bb {K}_1^F +\bb {K}_2^F$. This nomenclature comes from the $d=2$ 1-filling surface case, where there are only two solutions: $\bb{K}_1=\bb{K}_3, \bb{K}_2=\bb{K}_4$ or $\bb{K}_1=\bb{K}_4, \bb{K}_2=\bb{K}_3$. These two setups are equivalent to one another up to changing the Fermion order.
\subsection{1 Filling surface}
When there is only a single filling surface, the forward scattering is determined only by the solution to $\bb{K}_1=\bb{K}_3, \bb{K}_2=\bb{K}_4$. 
Including spins, there are explicitly 3 choices:
    \begin{align}
        F^1(\bb n_1,\bb n_2)=u(\bb K_2\sigma,\bb K_1\sigma,\bb K_2\sigma,\bb K_1\sigma),\label{eq:1FSF1}\\
        F^2(\bb n_1,\bb n_2)=u(\bb K_2\bar\sigma,\bb K_1\sigma,\bb K_2\bar\sigma,\bb K_1\sigma),\\
        F^3(\bb n_1,\bb n_2)=u(\bb K_2\sigma,\bb K_1\bar\sigma,\bb K_2\bar\sigma,\bb K_1\sigma).
        \label{eq:1FSF3}
    \end{align}
Due to the fact that any higher order terms in the Taylor expansion on $\omega$ and $k$ are irrelevant, we can freely choose the frequency and momentum deviations. For simplicity, we set all external legs to zero frequency and almost on the filling surface ($\omega=0,k=\epsilon\ll d\Lambda$). The tiny value of $\epsilon$ will be set equal to zero at the last step. We need to keep it during the calculation to make the running momentum $K$ and $K+Q$ different according to the requirement of the weak coupling expansion.

First, we consider the ZS diagram in Fig.\ref{feynmandiagram} given by the first integral in Eq.(\ref{eq:diags}). Since $Q=\epsilon\ll d\Lambda$, both $K$ and $K+Q$ lie on the same side of the filling surface for all eligible choices of $K$. As a result, the directional integral of $K$ is over the full range,
\begin{equation}
\begin{split}
    dF(\bb n_1,\bb n_2)=&\int \frac{d\bb n}{(2\pi)^{d-1}} F(\bb n_1,\bb n)F(\bb n,\bb n_2)\\
    &\int_{d\Lambda}\frac{dk}{2\pi}\int_{-\infty}^\infty \frac{d\omega}{2\pi} G(\omega,k)G(\omega,k+\epsilon).
\end{split}
\end{equation}
Here $F(\bb n_1,\bb n_3)$ represents the appropriate choice of $F$ that satisfies the momentum and spin conservation at each vertex.
The integral over $dk$ lies inside the thin shells to be integrated out. However, there are two such shells.  One of the shells lies inside the filling surface while the other is outside the filling surface. For the outer shell corresponding to $k\in[\Lambda-d\Lambda,\Lambda]>0$, the states belong to the 0-occupancy region, which means we can replace the Green function by
\beq
G(\omega,k)=\frac{1}{i\omega-v_Lk}.
\eeq
For the inner shell, corresponding to $k\in[-\Lambda,-\Lambda+d\Lambda]<0$, the states belong to the single-occupancy region, which means we can replace the Green function by
\beq
G(\omega,k)=\frac{1/2}{i\omega-v_Lk}+\frac{1/2}{i\omega-v_Lk-U}.
\eeq
The poles in the $\omega$ plane do not contribute if they lie on the same side of the real axis.  The only surviving contribution from ZS is thus
\begin{equation}
\begin{split}
    dF(&\bb n_1,\bb n_2)=2\int \frac{d\bb n}{(2\pi)^{d-1}} F(\bb n_1,\bb n)F(\bb n,\bb n_2)\\
    &\int_{-\Lambda}^{-\Lambda+d\Lambda}\frac{dk}{2\pi}\int_{-\infty}^\infty \frac{d\omega}{2\pi} \frac{1/2}{i\omega-v_Lk}\cdot\frac{1/2}{i\omega-v_Lk-U}.
\end{split}
\end{equation}
The integral over $\omega$ and $k$ gives
\begin{equation}
    \int_{d\Lambda}\frac{dk}{2\pi}\int_{-\infty}^\infty \frac{d\omega}{2\pi} \frac{1/2}{i\omega-v_Lk}\cdot\frac{1/2}{i\omega-v_Lk-U}=\frac{d\Lambda}{8\pi U}.
\end{equation}
As $U$ is strongly relevant, that is, $U'=s^{-2}U$, this contribution goes to zero much faster than $d\Lambda/\Lambda$. Thus the ZS diagram does not yield any contribution under the RG analysis.

Now consider the ZS' diagram. Due to the momentum transfer $Q$ of order $K_L$ at the left vertex, not only is the magnitude of the loop momentum restricted to lie within the shell being eliminated but also its angle is restricted to a range of order $d\Lambda/K_L$. 
This suppression in ZS' diagrams contributes to $d\Lambda^2/\Lambda K_L$. The $\beta$ function thus vanishes as we take the limit $d\Lambda/\Lambda\rightarrow0$. 

Finally, the same kinematic reasons used to establish that ZS' vanishes can be adopted to show that the BCS diagram also does not renormalize $F$ at one loop.  Hence, 
the coupling constants for the forward scattering do not flow in this order because $\beta(F)=0$.

\subsection{2 Filling surfaces}

\begin{figure}[h]
\centering
\includegraphics[scale=.3]{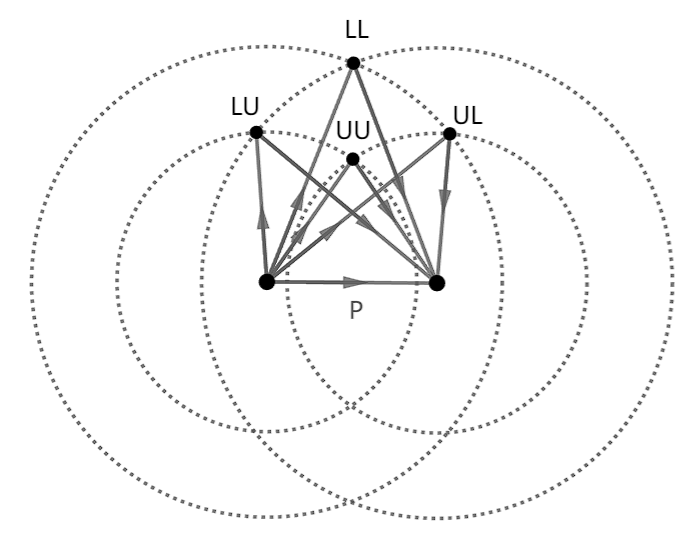}
\caption{The solutions to $\bb{P}=\bb{K}_1+\bb{K}_2$. The L-surface and U-surface are plotted in dotted circles. The 4 solutions are marked by the surfaces where the 2 momentums that sum up to $\bb{P}$ are.}
\label{2fs}
\end{figure}

When there are 2 filling surfaces, there are at most 4 independent solutions to $\bb{P}=\bb{K}_1+\bb{K}_2$ as shown in Fig.\ref{2fs}.
The combined equation $\bb{P}=\bb{K}_1+\bb{K}_2$ and $\bb{P}=\bb{K}_3+\bb{K}_4$ thus have $4\times4=16$ independent solutions.
There are still explicitly 3 spin configurations for each momentum solution. Thus the total number of marginal forward scattering is $16\times3=48$. We
will not enumerate them here since they remain un-
changed in the same way as we discussed in the case of a single-filling surface. The 1-loop correction contributes a term proportional to either $d\Lambda/U$ from the ZS diagrams or $d\Lambda/K_L$ for the ZS' and BCS diagrams. Once again, forward scatterings remain fixed under RG; that is, they do not flow under RG.

\section{Superconducting pairings at 1-loop}
The incoming and outgoing momenta each summing to zero $\bb{K}_1=-\bb{K}_2,\bb{K}_3=-\bb{K}_4$ constitute superconducting pairings. The 1-loop evolution of $V$s has a non-vanishing contribution even for a Fermi liquid and hence we expect an analogous contribution at the HK fixed point should one exist. We will analyze the single and two filling surfaces separately.

\subsection{1 Filling surface}
$\bb{K}_1$ and $\bb{K}_3$ can lie freely on the filling surface. For simplicity, we set all external legs to zero frequency and on the filling surface ($\omega=k=0$).
Including the spin-singlet and spin-triplet pairings, we have 2 choices,
    \begin{align}
        V^1(\bb n_1,\bb n_3)=u(-\bb K_3\sigma,\bb K_3\sigma,-\bb K_1\sigma,\bb K_1\sigma),\label{eq:1FSV1}\\
        V^2(\bb n_1,\bb n_3)=u(-\bb K_3\bar\sigma,\bb K_3\sigma,-\bb K_1\bar\sigma,\bb K_1\sigma).
        \label{eq:1FSV2}.
    \end{align}
These two spin configurations satisfy the antisymmetry for Fermions regardless of the angular structure of the pairing interaction.  The incoming momenta are equal and opposite on the filling surface. The ZS and ZS' diagrams are suppressed by $d\Lambda^2/\Lambda K_L$ and hence do not contribute for the same reason that ZS' and BCS diagrams did not contribute to the flow of forward scatterings. The BCS diagram now has a contribution since the running momentum in the loop can now freely point in any direction, regardless of the value of $K$. We rewrite Eq.(\ref{eq:diags}) as
\begin{equation}
\begin{split}
    dV^{1,2}(\bb n_1,\bb n_3)=&-\frac{1}{2}\int \frac{d\bb n}{(2\pi)^{d-1}} V^{1,2}(\bb n_1,\bb n)V^{1,2}(\bb n,\bb n_3)\\
    &\int_{d\Lambda}\frac{dk}{2\pi}\int_{-\infty}^\infty d\omega G(\omega,k)G(-\omega,k).
\end{split}
\label{eq:rgv}
\end{equation}
The integral over $dk$ also lies inside the two thin shells to be integrated out. These two thin shells yield contribute differently. For the outer shell, corresponding to $k\in[\Lambda-d\Lambda,\Lambda]>0$, the second line of Eq.(\ref{eq:rgv}) yields a finite value $\frac{d\Lambda}{4\pi\Lambda}$. For the inner shell, corresponding to $k\in[-\Lambda,-\Lambda+d\Lambda]<0$, the integral region thus gives a value that is reduced by a factor of a quarter, $\frac{d\Lambda}{16\pi\Lambda}$. In all, The renormalization group equation for $V^{1,2}$ is
\begin{equation}
\begin{split}
    \frac{dV(\bb n_1,\bb n_3)}{dt}=&-\frac{5}{32\pi v_L}\int \frac{d\bb n}{(2\pi)^{d-1}} V(\bb n_1,\bb n)V(\bb n,\bb n_3)\\
    \equiv&-\frac{5}{32\pi v_L}(V*V)(\bb n_1,\bb n_3),
\end{split}
\label{eq:betav}
\end{equation}
where $dt=|d\Lambda|/\Lambda$ is the RG transform step size, and $*$ defines the generalized convolution in $d$-dimensions.
This is the Cooper instability in the HK model. 
For the case of $d=2$, we can simplify this by going to momentum eigenfunctions
\beq
V_l=\int_0^{2\pi}\frac{d\theta}{2\pi}e^{il\theta}V(\theta),
\eeq
where $V(\theta)=V(\bb n_0,R_\theta\bb n_0)$ and $R_\theta$ is the rotation by degree $\theta$. We can obtain the $\beta$ function for each angular momentum $l$
\beq
\frac{dV_l}{dt}=-\frac{5}{32\pi v_L}V_l^2.
\eeq
The flow tells us that the couplings $V_l$ are marginally relevant if negative, and marginally irrelevant if positive. By integrating the flow, we obtain
\beq
V_l(t)=\frac{V_l(0)}{1+5tV_l(0)/32\pi v_L},
\eeq
which implies an instability at the energy scale $\Lambda_c=\Lambda_0e^{32\pi v_L/5V_0}$. The energy scale in a thermal system can be proportional to the temperature; thus we propose that the approximate transition temperature of this metallic state scales as $\Lambda_c$.

\subsection{2 Filling surfaces}

The two-filling surface case remains to be analyzed.  The electrons around different filling surfaces now have different contributions and can be grouped into 3 different categories (intra-L, intra-U, and inter-LU)
    \begin{align}
        V^1(\bb n_1,\bb n_3)=u(-\bb K_3^L\bar\sigma,\bb K_3^L\sigma,-\bb K_1^L\bar\sigma,\bb K_1^L\sigma),\label{eq:2FSV1}\\
        V^2(\bb n_1,\bb n_3)=u(-\bb K_3^U\bar\sigma,\bb K_3^U\sigma,-\bb K_1^U\bar\sigma,\bb K_1^U\sigma),\\
        V^3(\bb n_1,\bb n_3)=u(-\bb K_3^U\bar\sigma,\bb K_3^U\sigma,-\bb K_1^L\bar\sigma,\bb K_1^L\sigma),
        \label{eq:2FSV3}
    \end{align}
for the  spin-singlet configuration.   Here the superscripts represent the surface around which the momenta are  located. The spin-triplet processes have exactly the same RG structure and follow the same RG equations. We omit their definition and work with only the spin-singlet processes.

\begin{figure}[h]
\centering
\includegraphics[scale=.66]{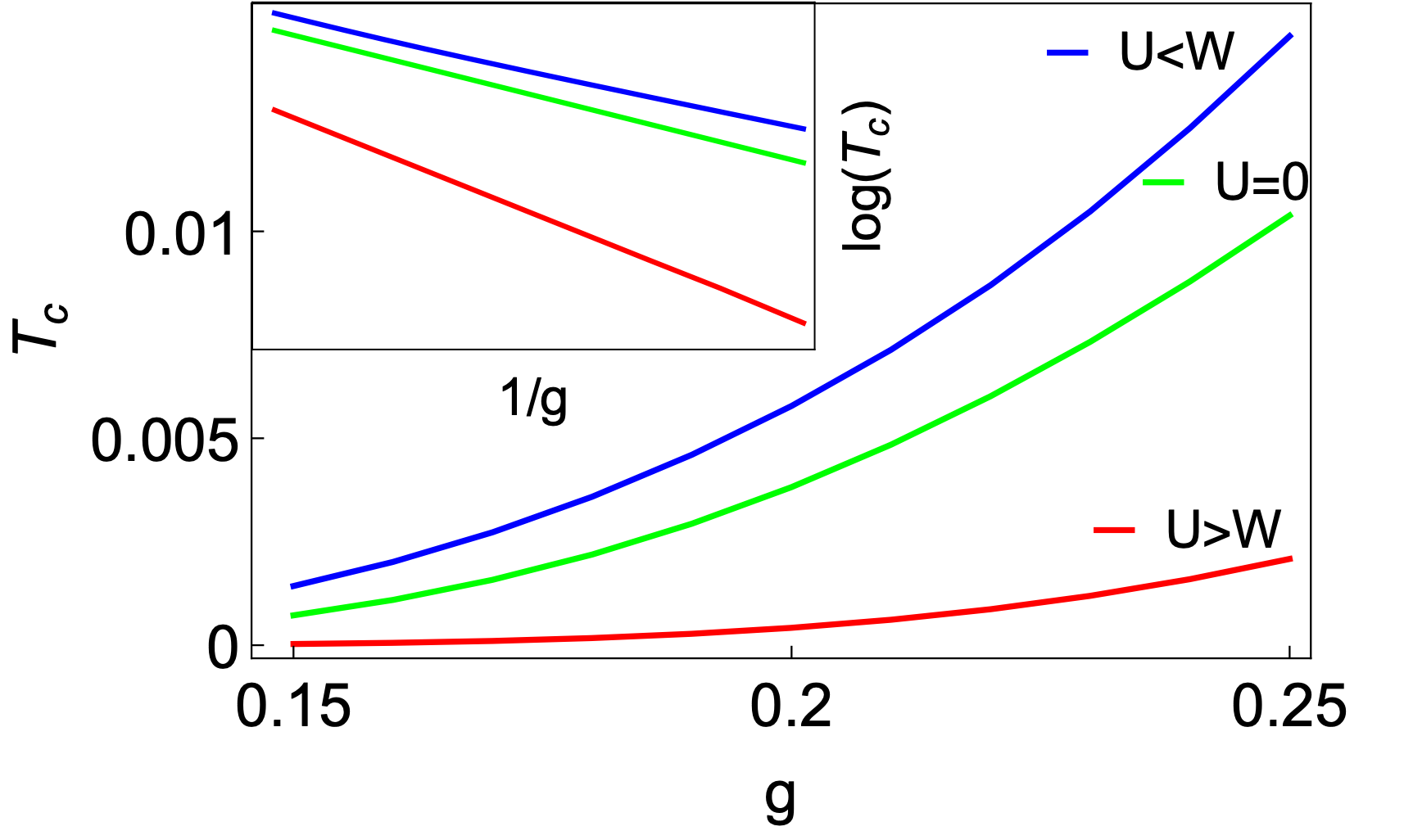}
\caption{The dependence of $T_c$ on superconducting pairing strength $g=-V$. The curves from top to bottom are: $U<W$ (2 filling surfaces), $U=0$ (Fermi liquid), and $U>W$ (1 filling surface). The insert plot shows the linear dependence of $\log(T_c)$ on $1/g$ and the slope of each curve is $0.8:1:1.6$. These $T_c$ exponents are consistent with the $\Lambda_c$ estimate.}
\label{Tc}
\end{figure}

The RG flow of the BCS couplings still follows Eq.(\ref{eq:rgv}). The running momentum can be chosen freely on both filling surfaces. The integral of the Green function around each filling surface follows the same process as Eq.(\ref{eq:betav}).   The renormalization group equations are
\begin{align}
    \frac{dV^1}{dt}=&-\frac{5}{32\pi}(\frac{V^1*V^1}{v_L}+\frac{V^{3\dagger}*V^3}{v_U}),\\
    \frac{dV^2}{dt}=&-\frac{5}{32\pi}(\frac{V^3*V^{3\dagger}}{v_L}+\frac{V^2*V^2}{v_U}),\\
    \frac{dV^3}{dt}=&-\frac{5}{32\pi}(\frac{V^1*V^3}{v_L}+\frac{V^3*V^2}{v_U}).
\end{align}
With the simplification, $V^1=V^2=V^3=V$, we simplify the $\beta$ function to a single equation,
\begin{equation}
    \frac{dV}{dt}=-\frac{5}{32\pi}(\frac{1}{v_L}+\frac{1}{v_U})V*V.
\end{equation}
For the case of $d=2$, the instability for attractive $V$ obtains at the energy scale 
\begin{equation}
\Lambda_c=\Lambda_0\exp\left({\frac{32\pi v_Lv_U}{5(v_L+v_U)V_0}}\right).
\end{equation}
The energy scale in a thermal system can be proportional to the temperature.
This increase in the  critical energy scale is consistent with the exactly calculated pair susceptibility\cite{HKsupercon}, which diverges at a higher temperature for a small value of $U$ compared with the Fermi liquid result. The comparison between different $T_c$ was shown in Figure(\ref{Tc}) 

Due to the fact that RG analysis only deals with perturbative interactions around the fixed point at each step, the first order transition into a superconductor at finite $U$\cite{HKsupercon} is absent.
\section{Final Remarks}

\begin{figure}[ht]
\includegraphics[width=\columnwidth]{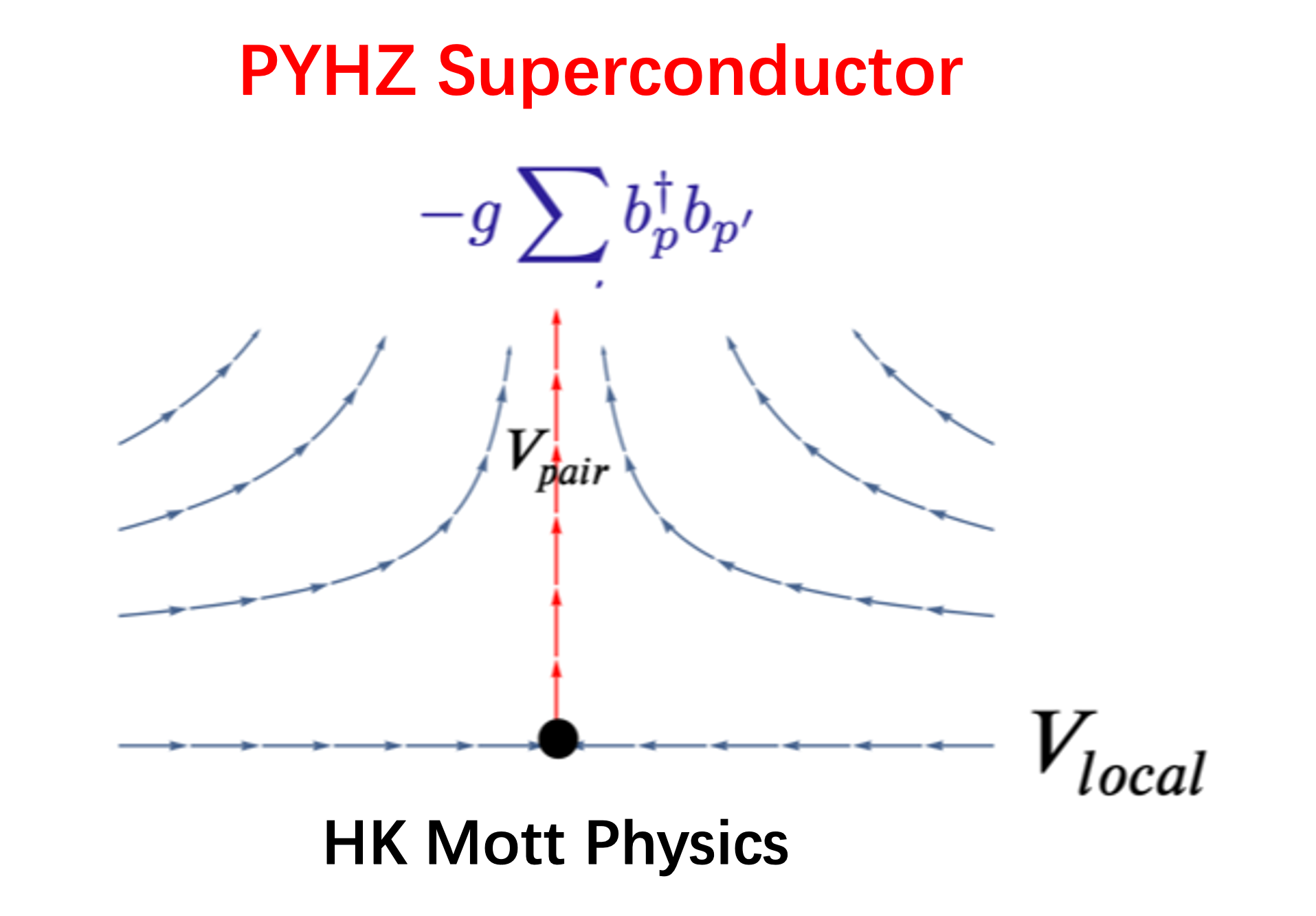}
\caption{Perturbative flow diagram for interactions in a doped HK Mott system.  Short-range interactions regardless of their sign do nothing.  Only pairing leads to flow to strong coupling and the ultimate destruction of the non-Fermi liquid HK metallic state and the onset of a superconducting state distinct from that of a BCS superconductor.  The nature of the superconducting state cannot be established based on perturbative arguments but requires the PYH theory\cite{nat1,HKsupercon}.}
\label{hkflow}
\end{figure}

We have shown that the analogies with Fermi liqiud theory noted in the introduction are borne out by a thorough RG analysis of the interactions that can contribute to the HK model.  For short-range repulsions, nothing flows away from HK, thereby establishing it as a stable fixed point depicted in Fig. (\ref{hkflow}).  While it is surprising that a new quartic fixed point exists, it is ultimately not surprising given that the momentum structure of the HK model is identical to that of Fermi liquid theory.  As a consequence, all of the interactions which flow into Fermi liquid theory also flow into this quartic fixed point.  
Once again as in Fermi liquid theory, superconductivity leads to flow away from the HK fixed point (see Fig.\ref{hkflow}).  Since Hubbard interactions are also local in real space, they also cannot perturb away from the HK fixed point in the metallic state.  Consequently, the simplicity of the HK model belies its true robustness and generality. 

\begin{figure}[h]
    \centering
    \includegraphics[scale=.5]{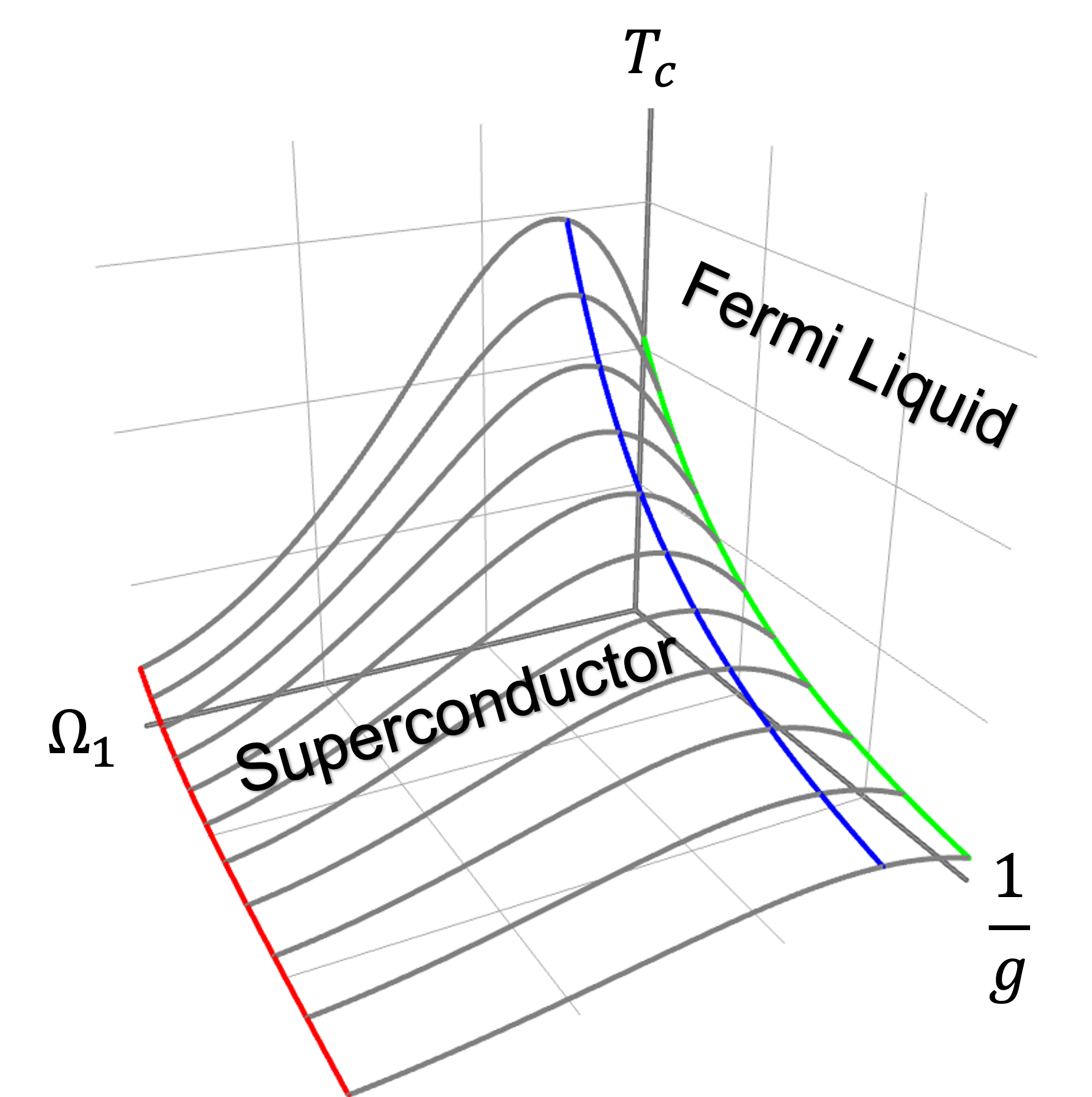}
    \caption{The general phase diagram of the superconducting instability in the HK fixed point. As long as the $\Omega_1$ region is non-zero, the system breaks the Fermi Liquid $Z_2$ symmetry and flows into HK. }
    \label{fig:3dphase}
\end{figure}

While the similarity of the momentum structure of the theory with that of Fermi liquid theory plays a key role in the stability of the fixed point, the relation to Hubbard is ultimately driven by the $Z_2$ symmetry breaking of the interaction in Eq. (\ref{eq1}).  As long as the interaction is repulsive, breaking the $Z_2$ symmetry noted by Anderson and Haldane\cite{Haldane} leads to single occupancy.
The phase diagram for the evolution of HK physics from a simple Fermi liquid can then be plotted as a function of the singly occupied region, $\Omega_1$,
as shown in Fig. (\ref{fig:3dphase}).  Fermi liquids live only in the region where $\Omega_1$ vanishes.  The transition line for superconductivity from a FL state is given by
the green line.  The entire region in the $\Omega_1-g$ plane represents non-BCS superconductivity and is governed by the quartic HK fixed point delineated here.  In addition to mediating non-Fermi liquid behvaiour, single occupancy in real or momentum space leads to degeneracy.  Degeneracy is a tell-tale sign of competing ordering tendencies that are well known to obtain in strongly correlated systems.  For example, the pure HK model has a divergent ferromagnetic susceptibility
(See Appendix). What the HK model does is separate the bifurcation of the spectral function into low and high energy branches per momentum state, the inherent physics of a Mott insulator, from ordering tendencies.  Such ordering is ultimately needed to free the HK model of the spurious ground-state degeneracy without generating flow away from the stable fixed point.  Hence, what the HK model ultimately does is offer a way of treating Mott's original conception of the gapped half-filled band.  Mottness sets the gap and ordering is secondary as borne out experimentally in all Mott systems ranging from the vanadates\cite{vo2} to the cuprates\cite{cuprate,cuprate1,cuprate2}.

\textbf{Acknowledgements} 
We thank the NSF DMR-2111379 for partial funding this work.
 
 \clearpage
\section*{Appendix}
\subsection{Detailed derivation of the Green function}
Here we briefly review the exact partition and Green function of the HK model at any filling. 
We start with the HK Hamiltonian,
\begin{equation}
    H_{HK}=\sum_k\left[\xi_k(n_{k\uparrow}+n_{k\downarrow})+Un_{k\uparrow}n_{k\downarrow}\right],
\end{equation}
where $\xi_k=\epsilon_k-\mu$.
For each momentum sector, the HK Hamiltonian is built out of commuting parts. Thus, the partition function factorizes, and for each momentum sector we have
\begin{equation}
    Z_k=1+2e^{-\beta\xi_k}+e^{-\beta(2\xi_k+U)}.
\end{equation}
The Heisenberg equations in imaginary time for Fermion
$c_{k\sigma}$ annihilation and the number operator $n_{k\sigma}=c^\dagger_{k\sigma}c_{k\sigma}$ are
\begin{align}
    \dot{c}_{k\sigma}&=\left[H,c_{k,\sigma}\right]=-(\xi_{k}+Un_{k\bar\sigma})c_{k\sigma},\\
    \dot{n}_{k\sigma}&=\left[H,n_{k,\sigma}\right]=0.
\end{align}
Thus we have the time evolution of the fermi operator,
\begin{equation}
    c_{k\sigma}(\tau)=e^{-(\xi_{k}+Un_{k\bar\sigma})\tau}c_{k\sigma}(0).
\end{equation}
The average particle number is
\begin{equation}
    \braket{n_{k\sigma}}=\frac{e^{-\beta\xi_k}+e^{-\beta(2\xi_k+U)}}{1+2e^{-\beta\xi_k}+e^{-\beta(2\xi_k+U)}}.
\end{equation}
The imaginary time Green function is
\begin{equation}
\begin{split}
    G_{k\sigma}(\tau)&=-\braket{c_{k\sigma}(\tau)c^\dagger_{k\sigma}(0)}\\
    &=-\mathrm{Tr}\left[e^{-\beta(H-F)}e^{-(\xi_{k}+Un_{k\bar\sigma})\tau}c_{k\sigma}(0)c^\dagger_{k\sigma}(0)\right]\\
    &=-\mathrm{Tr}\left[e^{\beta F}e^{-(\xi_{k}+Un_{k\bar\sigma})\tau}e^{-\beta H}(1-n_{k\sigma})\right]\\
    &=-e^{\beta F}\left[e^{-\xi_k\tau}+e^{-\beta\xi_k}e^{-(\xi_k+U)\tau}\right].
\end{split}
\label{eq:greenf}
\end{equation}
Performing the Fourier transform with the anti-periodic boundary condition $G(\tau+\beta)=-G(\tau)$ leads to
\begin{equation}
\begin{split}
    G_{k\sigma}(i\omega_n)&=\int_0^\beta G_{k\sigma}(\tau)e^{i\omega_n\tau}\\
    &=-e^{\beta F}\left[\frac{-e^{-\beta\xi_k}-1}{i\omega_n-\xi_k}+e^{-\beta\xi_k}\frac{-e^{-\beta(\xi_k+U)}-1}{i\omega_n-\xi_k-U}\right]\\
    &=\frac{1-\braket{n_{k\bar\sigma}}}{i\omega_n-\xi_k}+\frac{\braket{n_{k\bar\sigma}}}{i\omega_n-\xi_k-U}.
\end{split}
\end{equation}
This result is exact for any value of $\mu$ and $U$.

\subsection{Weak coupling Expansion and Wick's Theorem} \label{sec:weak-couple}
According to the Gell-Mann-Low formula, the $2n$-point function under interaction $V$ is given by
\begin{widetext}
\begin{equation}
\begin{split}
    &\braket{Tc_{k_1}^\dagger(\tau_1)\cdots c_{k_n}^\dagger(\tau_n) c_{k'_1}(\tau'_1)\cdots c_{k'_n}(\tau'_n)}_{I}\\
    &\quad=\frac{\braket{T\bar c_{k_1}(\tau_1)\cdots \bar c_{k_n}(\tau_n) c_{k'_1}(\tau'_1)\cdots c_{k'_n}(\tau'_n)\exp\left(-\int_0^\beta d\tau V\left[\bar c,c\right]\right)}_{HK}}{\braket{T\exp\left(-\int_0^\beta d\tau V\left[\bar c,c\right]\right)}_{HK}},
\end{split}
\label{eq:gellmann}
\end{equation}
\end{widetext}
where $\braket{}_{I}$ is the average using the full Hamiltonian $H_{HK}+H_I$ and $\braket{}_{HK}$ is the average under the HK Hamiltonian $H_{HK}$ only.

Since the HK path integral is decomposed into a series of products in the momentum space, the creation and annihilation operators from different momentum sectors can be safely factored out.  For example, suppose $k_1,k_2,\cdots,k_n$ are different from each other, and each corresponding annihilation and creation operator appears $l_1,l_2,\cdots,l_n$ times, then
\begin{widetext}
\begin{equation}
\begin{split}
    &\left\langle T\prod_{j=1}^{l_1} c_{k_1\sigma_1^j}^\dagger(\tau_1^{j})\cdots \prod_{j=1}^{l_n} c_{k_n\sigma_n^j}^\dagger(\tau_n^{j}) \prod_{j=1}^{l_1} c_{k_1\sigma_1^j}(\tau'_1{}^{j})\cdots \prod_{j=1}^{l_n} c_{k_n\sigma_n^j}(\tau'_n{}^{j})\right\rangle_{HK}\\
    &=(-1)^P\left\langle T\prod_{j=1}^{l_1} c_{k_1\sigma_1^j}^\dagger(\tau_1^{j})\prod_{j=1}^{l_1} c_{k_1\sigma_1^j}(\tau'_1{}^{j})\right\rangle_{HK}\cdots\left\langle T\prod_{j=1}^{l_n} c_{k_n\sigma_n^j}^\dagger(\tau_n^{j})\prod_{j=1}^{l_n} c_{k_n\sigma_n^j}(\tau'_n{}   ^{j})\right\rangle_{HK},
\end{split}
\end{equation}
\end{widetext}
where $P$ is the times of permutation performed so as to separate the fermion operators.

The deviation from Wick's theorem could be observed on certain multi-point correlation functions(e.g. 4-point function) for which all the momenta of the fermion operators are identical. This contribution, however, is thermodynamically suppressed in our calculation. Thus we employ Feynman diagram rules and use the exact Green function to calculate the correlation functions.

\section{Magnetic instability}

The singly occupied region of the HK model at finite $U$ in the  HK model leads to a singly occupied region which introduces spin  degeneracy for each momentum sector. This ground state degeneracy is extensive with the system size and thus considered to be unphysical although a singly occupied region generally exists in Mott systems with the traditional Hubbard model. The solution to the degeneracy in the HK fixed point is ordering such as superconductivity, ferromagnetism or a spin-density wave. The superconducting instability has been discussed in the main body and we provide here an analysis of a possible magnetic ordering of the HK model.

With an external magnetic field $B$, the partition function now reads,
\begin{equation}
    Z(B)=\sum_k\left(1+e^{-\beta(\xi_k-\mu B)}+e^{-\beta(\xi_k+\mu B)}+e^{-\beta(2\xi_k+U)}\right).
\end{equation}
The magnetic susceptibility is achieved by a double derivative of the partition function,
\begin{equation}
\begin{split}
    \chi&=\left.-\frac{1}{\beta Z(B)}\frac{\partial^2 Z}{\partial B^2}\right|_{B=0}\\
    &=\left.\frac{\mu^2\beta}{Z(B)}\sum_k\left(e^{-\beta(\xi_k-\mu B)}+e^{-\beta(\xi_k+\mu B)}\right)\right|_{B=0}\\
    &=\mu^2\Omega_1\beta,
\end{split}
\end{equation}
where $\Omega_1$ is the extent of singly occupied region. The susceptibility diverges as temperature goes to zero, signaling a ferromagnetic phase transition at $T=0$, thus offering an avenue for resolving the ground state degeneracy.  As pointed out in the text, superconductivity also lifts the degeneracy and leads to flow away from the HK fixed point.


 \bibliography{rgbib}

\end{document}